\documentclass[]{spie}  

\usepackage{amsmath,amsfonts,amssymb}
\usepackage{graphicx}
\usepackage{subfigure}
\usepackage[colorlinks=true, allcolors=blue]{hyperref}
\usepackage{svg}
\usepackage{gensymb}
\usepackage{amssymb}
\usepackage{wasysym}
\usepackage{textcomp}
\usepackage{tcolorbox}

\title{SCALES-DRP : A Data Reduction Pipeline for an
Upcoming Keck Thermal Infrared Spectrograph}

\author[a]{Athira Unni}
\author[a]{Steph Sallum}
\author[b]{Peyton Benac}
\author[b]{Michael P. Fitzgerald}
\author[c]{Max Brodheim}
\author[c]{Rosalie McGurk}
\author[a]{Andy Skemer}
\author[a]{William T. S. Deich}
\affil[a]{Department of Astronomy and Astrophysics, University of California, Santa Cruz, USA}
\affil[b]{Division of Physical Sciences, University of California Los Angeles, LA, USA}
\affil[c]{W. M. Keck Observatory, Hawai’i, USA}
\authorinfo{Further author information: (Send correspondence to Athira)\\Athira Unni: E-mail: athira.exo@gmail.com \\Steph Sallum: E-mail: ssallum@ucsc.edu}

\begin{document} 
\maketitle

\begin{abstract}
 We present the end-to-end data reduction pipeline for SCALES (Slicer Combined with Array of Lenslets for Exoplanet Spectroscopy), the upcoming thermal-infrared, diffraction-limited imager, and low and medium-resolution integral field spectrograph (IFS) for the Keck II telescope. The pipeline constructs a ramp from a set of reads and performs optimal extraction and $\chi^2$ extraction to reconstruct the 3D IFS datacube. To perform spectral extraction, wavelength calibration, and sky subtraction, the pipeline utilizes rectification matrices produced using position-dependent lenslet point spread functions (PSFs) derived from calibration exposures. The extracted 3D data cubes provide intensity values along with their corresponding uncertainties for each spatial and spectral measurement. The SCALES pipeline is under active development, implemented in {\ttfamily Python} within the Keck data reduction framework, and is openly available on \hyperlink{https://github.com/scalessim/SCALES-DRP}{{\ttfamily GitHub}} along with dedicated documentation.
\end{abstract}


\section{INTRODUCTION} \label{sec:intro}  
SCALES (Slicer Combined with Array of Lenslets for Exoplanet Spectroscopy) is a next-generation thermal infrared coronagraphic imager and integral field spectrograph (IFS) for the W. M. Keck Observatory. Designed to be coupled with the upcoming High order Advanced Keck Adaptive optics (HAKA) system \cite{Scott_haka}, SCALES is projected for first light in early 2026. SCALES will detect and characterize a wide range of exoplanets, while also enabling observations of protoplanetary disks, Solar System targets such as Jupiter’s moon Io, as well as Galactic and extra-galactic sources \cite{sallum_2023}.

\begin{figure}[!htbp]
\centering
\rotatebox{0}{\includegraphics[height=6cm]{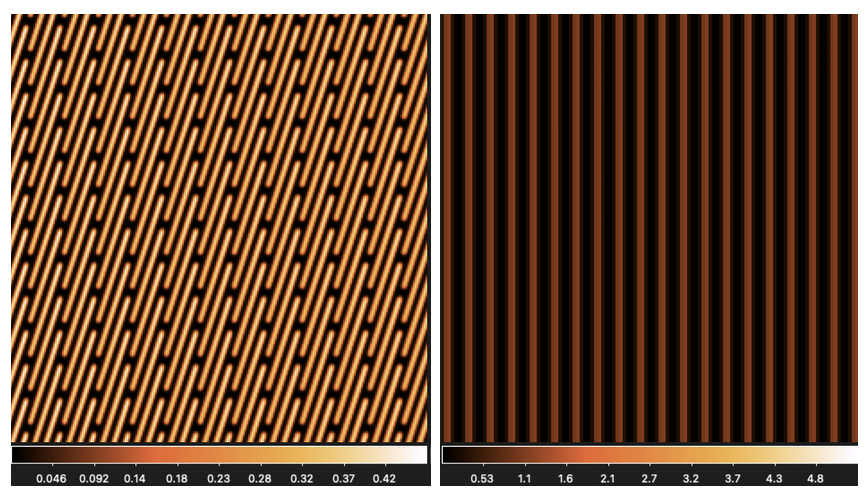}}
\caption[detectorflat] 
    { \label{fig:detectorflat} 
   A simulated IFS lenslet flat for SCALES in low-resolution (left) and medium-resolution (right) K-band modes. The microspectra are 54 pixels long in the low-resolution mode and 1900 pixels long in the medium-resolution mode, with adjacent microspectra separated by 6 pixels.} 
\end{figure}

The instrument will offer three distinct observing modes, including two IFS modes that cover the $2-5~\mu$m wavelength range. The low-resolution mode (R $\approx$ 35 - 250) utilizes a  silicon lenslet array, typically ranging from 102$-$105 $\times$ 92-100 lenslets, with a set of six prisms used to disperse the light and sample a $2.16^{\prime\prime} \times 2.16^{\prime\prime}$ field of view \cite{surya_2024}. A medium-resolution mode (R $\approx$ 2000 - 6500) uses an image slicer to reformat a $0.34^{\prime\prime} \times 0.36^{\prime\prime}$ sub-array of lenslets ($17 \times 18$) into a long pseudo-slit, which is then dispersed by diffraction gratings. The lengths of individual microspectra are approximately 54 pixels in the low-resolution mode and approximately 1900 pixels in the medium-resolution mode, with a separation of 6 pixels between the adjacent microspectra. Figure \ref{fig:detectorflat} shows a simulated IFS lenslet flat in K-band for both low-resolution and medium-resolution modes, illustrating the arrangement of the microspectra. Additionally, a dedicated imaging channel offers a wide $12.3^{\prime\prime} \times 12.3^{\prime\prime}$ field of view with a suite of 16 selectable filters spanning the $1–5~\mu$m wavelength range \cite{Banyal_2022}. 

Both the spectrograph and imaging channels will employ \textit{JWST}/NIRSpec spare Teledyne HAWAII-2RG (H2RG) detectors with a 5.3 $\mu$m cutoff \cite{Rauscher_2014}. The SCALES optical design is entirely reflective except a few transmissive components include the cryostat window, coronagraph masks, the lenslet array, prisms, and filters. The entire optical train is housed within a cryogenic environment to minimize thermal background noise \cite{Deno_2020, Skemer_2022, Kupke_2022}.

While building on the heritage of successful high-contrast instruments such as GPI \cite{Macintosh_2018}, SPHERE \cite{Beuzit_2006}, and CHARIS \cite{Groff_2015}, the unique capability of SCALES is its observation in the thermal infrared. This spectral window, combined with the extreme wavefront correction from HAKA, will push the frontiers of exoplanet science by enabling the direct detection and atmospheric characterization of cool, self-luminous gas giants beyond the reach of current ground-based facilities.

\section{OVERVIEW OF SCALES-DRP}
The final science product of an integral field spectroscopic (IFS) observation is a data cube, a three-dimensional array with two spatial axes corresponding to the lenslet grid and one spectral axis corresponding to wavelength. Each spatial element, or spaxel, contains a fully sampled spectrum of the source at that location. The SCALES data reduction pipeline (SCALES-DRP) reconstructs these 3D IFS data cubes from the raw detector reads, which contain microspectra from all lenslets, using two extraction algorithms: optimal extraction \cite{Horne_1986} and $\chi^{2}$ extraction \cite{Brandt_2017}. The DRP consists of three main modules: (1) a calibration module, (2) a quicklook module, and (3) a science grade module.

\begin{figure} [ht]
\begin{center}
\begin{tabular}{c}  
\includegraphics[height=2.4cm]{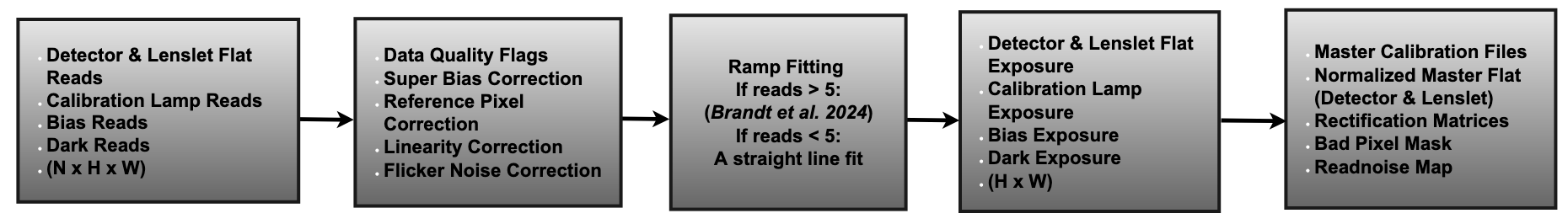}
\end{tabular}
\end{center}
\caption[flowchartcalib] 
    { \label{fig:flowchartcalib} 
    The workflow of the calibration module, starting from the input calibration raw data cube of shape $N \times H \times W$ (where $N$ is the number of reads), to the final outputs for both the IFS and imaging channels.}  
\end{figure}

The calibration module forms the core of SCALES-DRP, automatically processing daily afternoon calibration data required for science reductions. The monochromator data will be taken more intermittently as it is expected to remain stable over much longer timescales. The calibration module generates rectification matrices (RM) for optimal extraction, $\chi^2$ extraction, and wavelength calibration using a series of monochromatic calibration lamp exposures. The rectification matrices encode the information of each lenslet PSF (or psflet) across all wavelengths. This module also produces a bad pixel mask (BPM) using a series of flat and dark exposures, estimates the read-noise map from a set of bias exposures, and generates the master calibration files as well. Currently, the full calibration routine takes on the order of minutes to execute on a Mac M3 (128GB) laptop. A detailed explanation of the calibration module is given in Section \ref{calib:module}. Figure \ref{fig:flowchartcalib} shows the workflow of the calibration module.

A quicklook module is used for real-time analysis during observations and performs several key steps: reference pixel correction, linearity correction, and ramp fitting using a first-order polynomial (see Section \ref{ramp}), followed by detector and lenslet flat-fielding, as well as bad pixel correction (see Section \ref{bpm}). The final 3D IFS data cube is subsequently produced using a weighted optimal extraction, comprising two spatial dimensions and one spectral dimension (see Section \ref{spectral_extraction}). The quicklook module is designed for automatic rapid execution, which allows users to interactively visualize individual exposures and the final 3D IFS data cubes using a GUI during the observation itself.

\begin{figure}[!htbp]
\centering
\rotatebox{90}{\includegraphics[height=6.5cm]{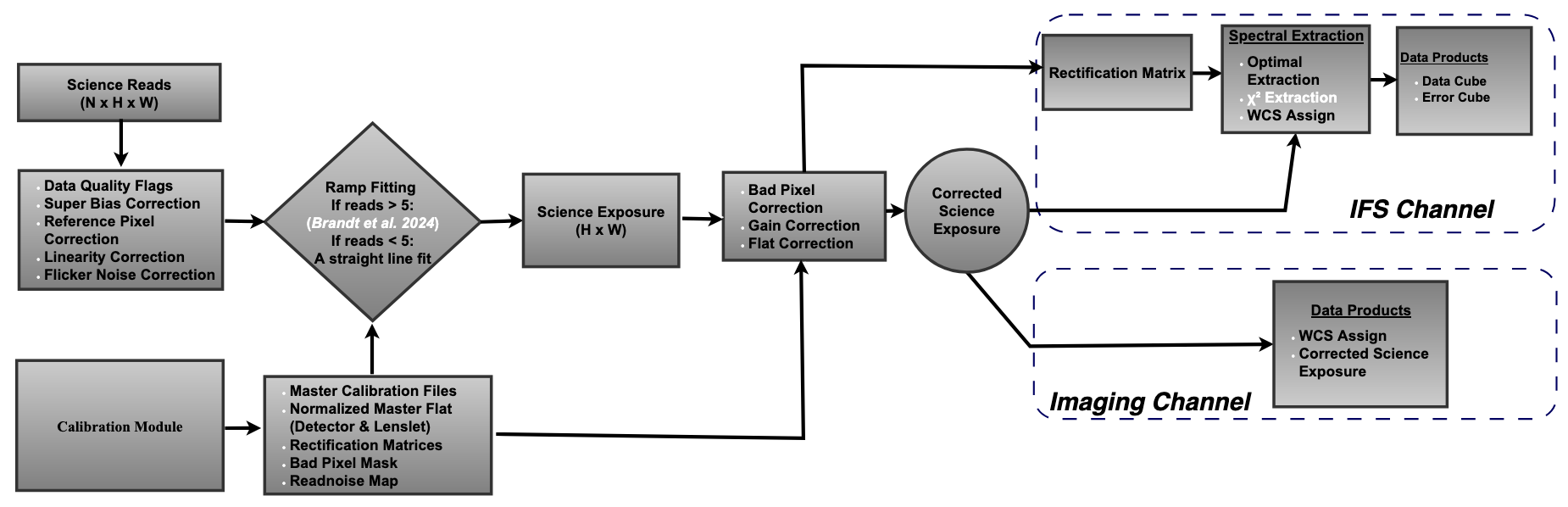}}
\caption[flowchartcalib] 
    { \label{fig:flowchartscience} 
    Workflow of the science module from raw detector reads to the final outputs for both the IFS and imaging channels. The quicklook module follows the same workflow but with faster execution time. Additional steps unique to the science module relative to the quicklook module are highlighted in white font. } 
\end{figure}

And finally, the science module performs detailed detector corrections, ramp fitting, and spectral extraction of the science reads, using outputs from the calibration module to generate the wavelength-calibrated 3D IFS datacube. The final 3D data cube output of the pipeline has been tested and validated using \href{https://github.com/scalessim/}{\ttfamily scalessim} \cite{Briesemeister_2020}, the SCALES simulator. The SCALES-DRP is parallelized, implemented in {\ttfamily Python} within the Keck-DRP framework, and openly available on \hyperlink{https://github.com/scalessim/SCALES-DRP}{{\ttfamily GitHub}} along with dedicated documentation. Figure \ref{fig:flowchartscience} shows the overall flowchart of the SCALES-DRP science module. 

\section{CALIBRATION MODULE} \label{calib:module}
The calibration module is the core of SCALES-DRP. It processes dark, bias, detector flat, and lenslet flat calibration files on a daily basis, followed by the afternoon calibration procedures at the observatory, and generates all the files required for the science module to process. Individual calibration exposures are created following reference pixel correction, 1/f correction, non-linearity correction, and a ramp fitting steps explained in Section \ref{ramp}. The calibration module generates rectification matrices using the master monochromatic calibration lamp exposures, a read-noise map from a set of bias exposures, and a bad pixel mask (BPM) followed by a BPM correction to all the calibration master files created. A five sigma-clipped mean is used to estimate the master bias, master calibration lamps, master dark, normalized master detector flat, and normalized master lenslet flat exposures (Figure \ref{fig:flowchartcalib}).

\subsection{Rectification Matrix}
\begin{figure} [ht]
\begin{center}
\begin{tabular}{c}  
\includegraphics[height=6cm]{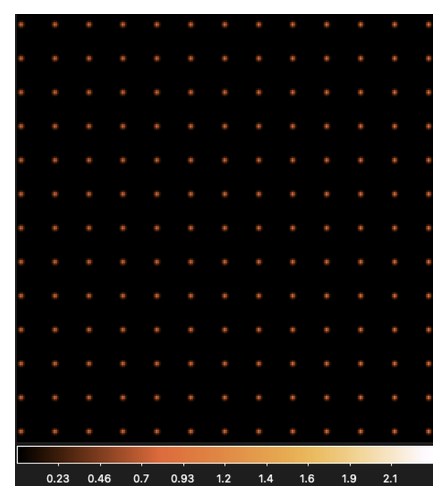}
\end{tabular}
\end{center}
\caption[calibimg] 
    { \label{fig:calibimg} 
    A simulated low-resolution calibration exposure at $2.20 \mu$m using {\ttfamily scalessim}. The Figure shows a zoomed view of a subset of the $108 \times 108$ lenslets.}  
\end{figure}

The rectification matrix (RM) is constructed empirically from a series of monochromatic calibration lamp exposures that densely sample the instrument's $2-5  \mu$m wavelength range. The SCALES calibration unit includes a monochromator capable of producing selectable, narrowband illumination at any central wavelength between $1-5 \mu$m (Lach, M. et al. in paper $\#13627-69$ in these proceedings). Each monochromatic lamp exposure produces a 2D image containing the point-spread functions (PSFs) from all lenslets, referred to as psflets. The number of wavelength bins in each observing mode is determined by the spectral resolution required for that mode. Figure \ref{fig:calibimg} is an example of a simulated low-resolution 2D calibration lamp exposure at $\lambda = 2.0 \mu$m using {\ttfamily scalessim}. 

The RM is a linear operator that forms the core of the instrumental model, mapping individual psflets across all wavelengths. This mapping is inherently sparse, since each monochromatic PSF is spatially localized and illuminates approximately a $3 \times 3$ box of detector pixels. The RM has dimensions $4194304,  X_{\rm lenslet} \times Y_{\rm lenslet} \times \lambda_{\rm bin}$, where 4194304 corresponds to the total number of pixels in the $2048 \times 2048$ detector, and $X_{\rm lenslet}$ and $Y_{\rm lenslet}$ represent the number of lenslets in the $x$ and $y$ directions, respectively (approximately $108 \times 108$ for low resolution and $17 \times 18$ for medium resolution).

Each column in the rectification matrix represents the weight of each pixel at each wavelength, measured from a calibration unit dataset. As a forward model, the RM transforms a physically meaningful 3D data cube ($A_{cube}$) into its corresponding 2D raw image ($d_{sim}$) via the linear operation $d_{sim} = R * A_{cube}$. Conversely, spectral extraction is formulated as the inverse problem: recovering the optimal three-dimensional data cube ($A_{\rm cube}$) that, when projected through the RM, best reproduces the observed 2D image (see Section \ref{spectral_extraction}). A similar RM-based approach has been employed by the OSIRIS integral field spectrograph at Keck II \cite{Lyke_2017, Lockhart_2019}. The calibration module further supports interpolation of the RM with sub-pixel accuracy, enabling corrections for small spatial or wavelength shifts between calibration unit exposures and science observations.

\subsection{Read-Noise Map}
\begin{figure} [ht]
\begin{center}
\begin{tabular}{c}  
\includegraphics[height=7cm]{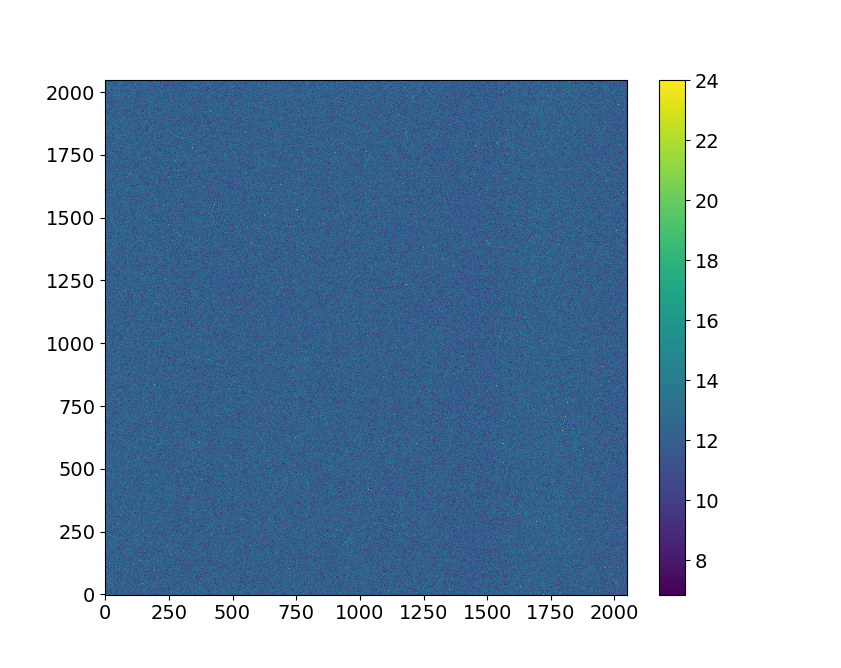}
\end{tabular}
\end{center}
\caption[readnoise] 
    { \label{fig:readnoise} 
    simulated read-noise map of a typical H2RG detector using the {\ttfamily Noise Generator}.}  
\end{figure}

The read-noise map is constructed from a sequence of bias exposures. A five sigma-clipped median bias frame is first derived and subtracted from individual exposures to account for the mean bias level. The read-noise map is then defined as the five sigma-clipped median of the resulting residual frames. As we have no proper lab bias exposures yet, for pipeline development we used a simulated read-noise map for an H2RG detector with a fundamental Gaussian noise floor, $1/f$ noise, a ``picture frame" effect with higher noise at the edges, alternative column noise (ACN), residual bias drifts, and a population of randomly distributed noisy pixels using the {\ttfamily Noise Generator} \cite{Rauscher_2015}. The read-noise map plays a critical role in ramp fitting, where it is used to construct the temporal covariance matrix for estimating correlated noise (Section \ref{ramp}). It is also essential for both $\chi^2$ and the optimal extraction, providing the proper pixel weighting required for accurate spectral reconstruction (Section \ref{spectral_extraction}). Figure \ref{fig:readnoise} shows the simulated read-noise map used for all the analysis presented in the paper.
\subsection{Bad Pixel Mask Generation} \label{bpm}
A bad pixel mask (BPM) is generated by identifying pixels that behave abnormally in either time or space, using stacks of dark and flat-field exposures. Unstable pixels are found using a temporal criterion: any pixel whose signal fluctuates by more than five sigma across the stack of exposures is flagged. Static hot or cold pixels are found using a spatial criterion: in each individual frame, a pixel is flagged if it deviates by more than five sigma from the median value of its local neighbors, calculated within a 5$\times$5 kernel. 

\begin{figure} [ht]
\begin{center}
\begin{tabular}{cc}  
\includegraphics[height=7cm]{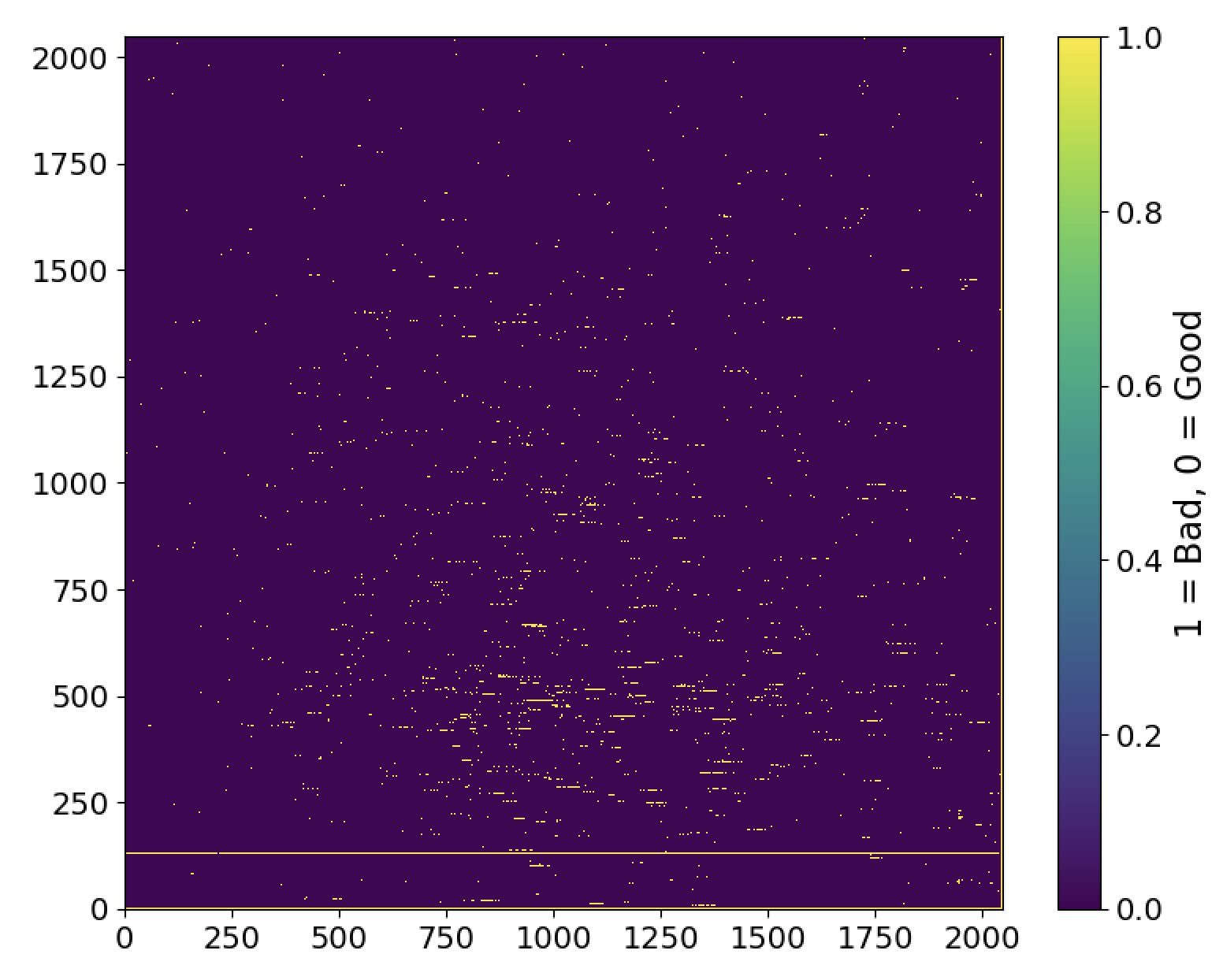}
\includegraphics[height=7cm]{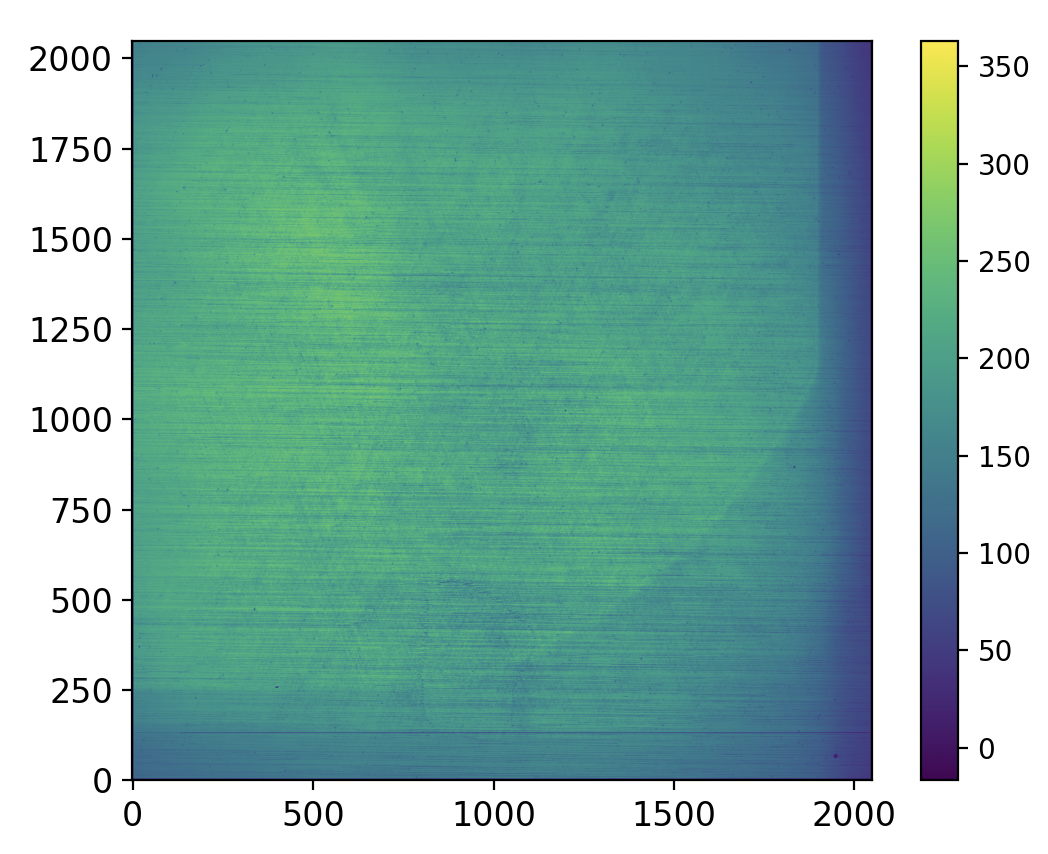}
\end{tabular}
\end{center}
\caption[bpm] 
    { \label{fig:bpm} 
    The bad pixel mask (BPM) derived from a cube of SCALES imager detector dark exposures and flat-field exposures using temporal and spatial cutoffs. Currently 1.58\% pixels are flagged as bad for the imager detector (left). The SCALES imager detector exhibits intrinsic cross-hatching in the HgCdTe crystal, which initially leads to the flagging of some pixels. This structure is effectively flattened by flat-fielding, and the bad pixel correction procedure is being refined to exclude pixels flagged solely due to the cross-hatching pattern. An imager dark exposure showing prominent horizontal cross-hatching patterns is also displayed (right).}  
\end{figure}

The final BPM combines all pixels flagged by either method, which currently totals 1.58\% of the imager detector array which is 0.23\% higher than the previous analysis by Rauscher et al.\cite{Rauscher_2014}. The SCALES imager detector exhibits a higher-than-average level of cross-hatching, an intrinsic feature of the HgCdTe crystal. As shown in Figure \ref{fig:bpm}, this effect initially leads to the flagging of some pixels affected by the cross-hatch pattern; however, the structure is effectively corrected through a detector flat-fielding. We are therefore in the process of modifying the bad pixel correction procedure to exclude pixels flagged solely due to the cross-hatch pattern and are further characterizing the BPM for the IFS detector. Further details on the bad pixel correction procedure are provided in Section \ref{ramp}.

\section{FROM READS TO RAMP IMAGES} \label{ramp}
\begin{figure} [ht]
\begin{center}
\begin{tabular}{c}  
\includegraphics[height=6cm]{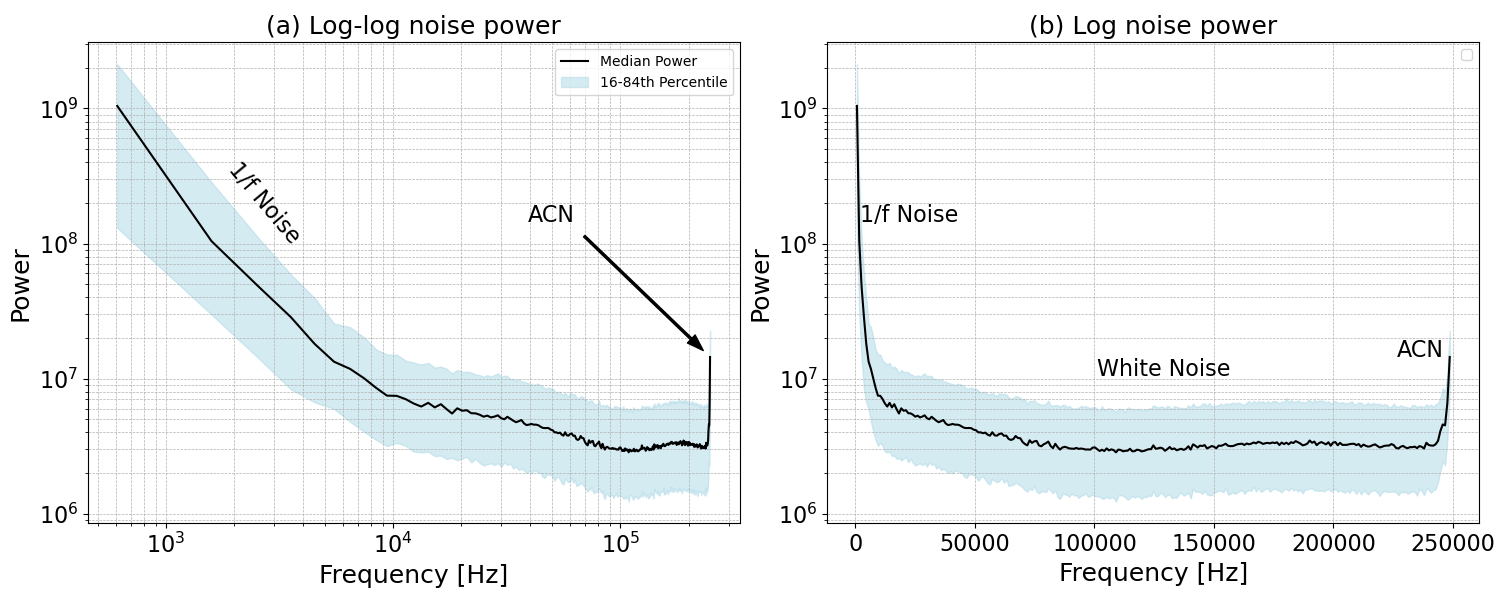}
\end{tabular}
\end{center}
\caption[flowchartcalib] 
    { \label{fig:read_median_fft} 
    The median of a fast fourier transform of a single SCALES read. When displayed on an (a) log-log scale, the 1/f low-frequency noise is obvious (the slope in the low frequency regime). In Figure (b), a linear-log scale to better show alternative column noise (ACN). ACN manifests as a strong feature at 7.3 MHz. The flat part of the power spectrum is roughly white noise.}  
\end{figure}

SCALES uses two $2048 \times 2048$ Teledyne H2RG detectors composed of HgCdTe. We read out the detectors with a Teledyne Imaging SIDECAR ASIC followed by an AstroBlank/Markury Scientific MACIE controller card, operating at a speed sufficient to avoid saturation during ground-based observations (Benac et al. in paper $\#13627-67$ in these proceedings). These detectors are sensitive from 0.6 to 5 $\mu$m and feature an architecture that divides the array into a central 2040×2040 area of light-sensitive pixels and a four-pixel-wide border of reference pixels. Although not exposed to light, these reference pixels electronically mimic the behavior of active pixels, allowing the correction of instrumental noise. They are used to track slowly changing offsets caused by drifting internal voltages, which addresses effects such as frame-to-frame bias shifts, even/odd column patterns, 1/f noise or flicker noise, and offsets between the four amplifier readout channels. A Fourier noise power spectrum is a powerful way of presenting these different noise sources (Figure \ref{fig:read_median_fft}). 

The input to SCALES-DRP is a data cube consisting of individual reads with dimensions (N $\times$ H $\times$ W). The pipeline applies several detector-level corrections to these reads: alternating column noise correction using the top and bottom reference pixels, 1/f noise correction using the left and right reference pixels, a pixel-level linearity correction, and finally ramp fitting with jump detection. These steps produce a ramp image, which serves as the basis for subsequent processing and are described in detail in the following subsections.

\subsection{Reference Pixel Correction}
\begin{figure} [ht]
\begin{center}
\begin{tabular}{c}  
\includegraphics[width=\textwidth]{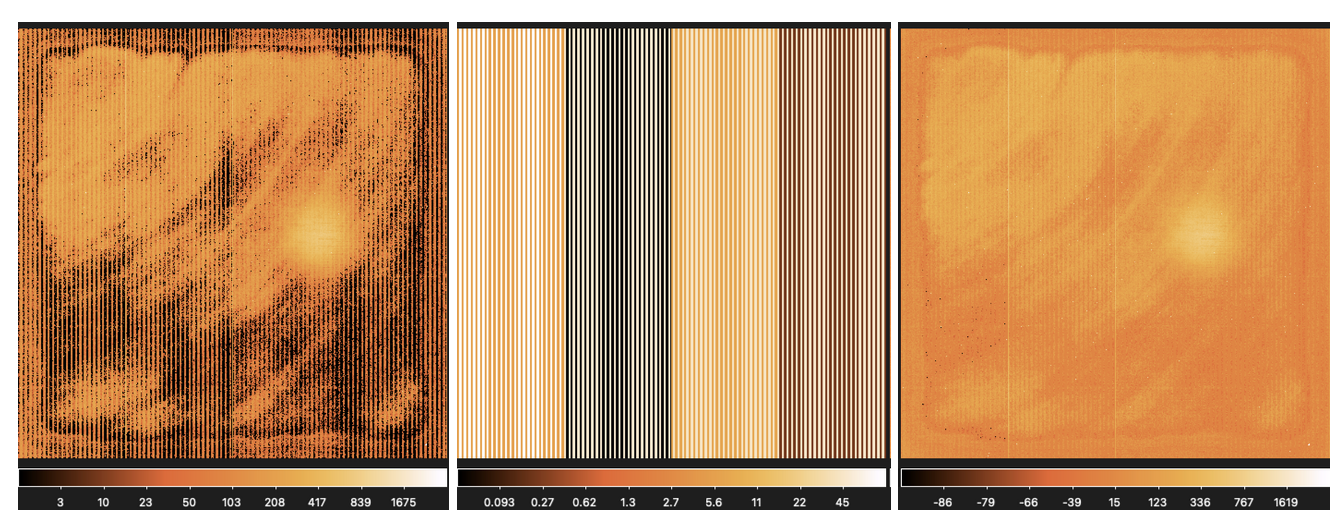}
\end{tabular}
\end{center}
\caption[refpix] 
    { \label{fig:refpix} 
    A raw read from the detector illuminated by a source (left), the alternative column noise estimated using the top and bottom reference pixels (middle), and the same read after the reference pixel correction (right).}  
\end{figure}

A reference pixel correction is applied to remove the offset between all four channels using the top and bottom reference rows. For that, a five-sigma-clipped mean value is estimated for even and odd reference columns separately, then this mean value is subtracted from the corresponding light sensitive column of pixels on the detector for each channel. Figure \ref{fig:refpix} shows a single read before and after the reference pixel correction.  

\subsection{1/f Correction}
A correction for flicker noise is implemented using the four columns of reference pixels located on the left and right sides of the detector. The core of this procedure is an optimal filtering algorithm, adapted from Kosarev \& Pantos \cite{Kosarev_1983,Rauscher_2015}, designed to isolate the low-frequency noise component from the reference pixel signal. This algorithm operates in the frequency domain to effectively distinguish the 1/f noise from the high-frequency random noise floor.

\begin{figure} [ht]
\begin{center}
\begin{tabular}{c}  
\includegraphics[width=\textwidth]{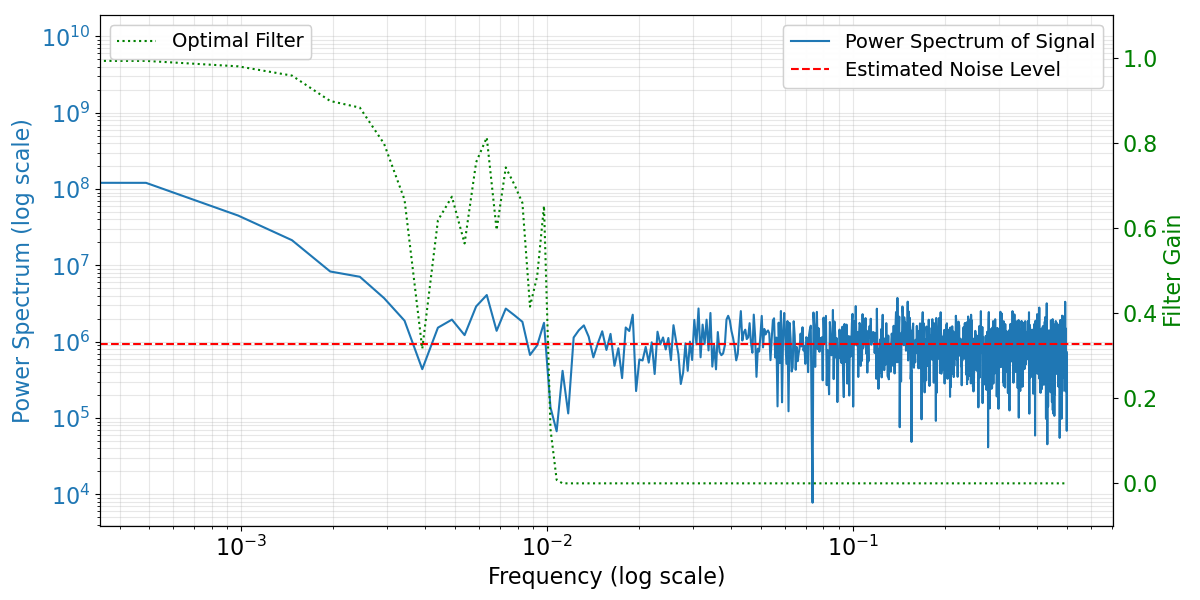}
\end{tabular}
\end{center}
\caption[fftimg1] 
    { \label{fig:fftimg1} 
    The power spectrum of the vertical reference pixels and the optimal filter profile used to extract the 1/f noise. The horizontal dashed line shows the mean value of the white noise at the higher frequencies. The high power region at the lower frequencies is the 1/f noise.}  
\end{figure}

The process begins by generating a one-dimensional noise signal by calculating the five-sigma-clipped mean of the eight available reference pixels. This signal is then transformed into the frequency domain via a Fast Fourier Transform (FFT). The subsequent filtering is based on the signal's own spectral properties. First, the algorithm establishes a white noise baseline by computing the average power in the upper quartile of the frequency spectrum, where random noise is assumed to be dominant. A frequency cutoff, J0, is identified where the signal's power spectrum first drops below this noise level. This cutoff separates the low-frequency domain, where 1/f noise is prominent, from the high-frequency domain dominated by random noise. Figure \ref{fig:fftimg1} illustrates the power spectrum of the reference signal of the imager detector, clearly showing the distinct flicker noise component at lower frequencies and the random noise floor at higher frequencies.

To model the 1/f noise component, a linear fit is applied to the logarithm of the power spectrum for all frequencies below J0. This model is then extrapolated to higher frequencies to estimate the underlying signal power where it is otherwise obscured by white noise. Following the principles of a Wiener filter, an optimal filter is constructed where the gain at each frequency is determined by the ratio of the modeled signal power to the total power (i.e., modeled signal plus the measured white noise baseline).

\begin{figure} [ht]
\begin{center}
\begin{tabular}{c}  
\includegraphics[width=\textwidth]{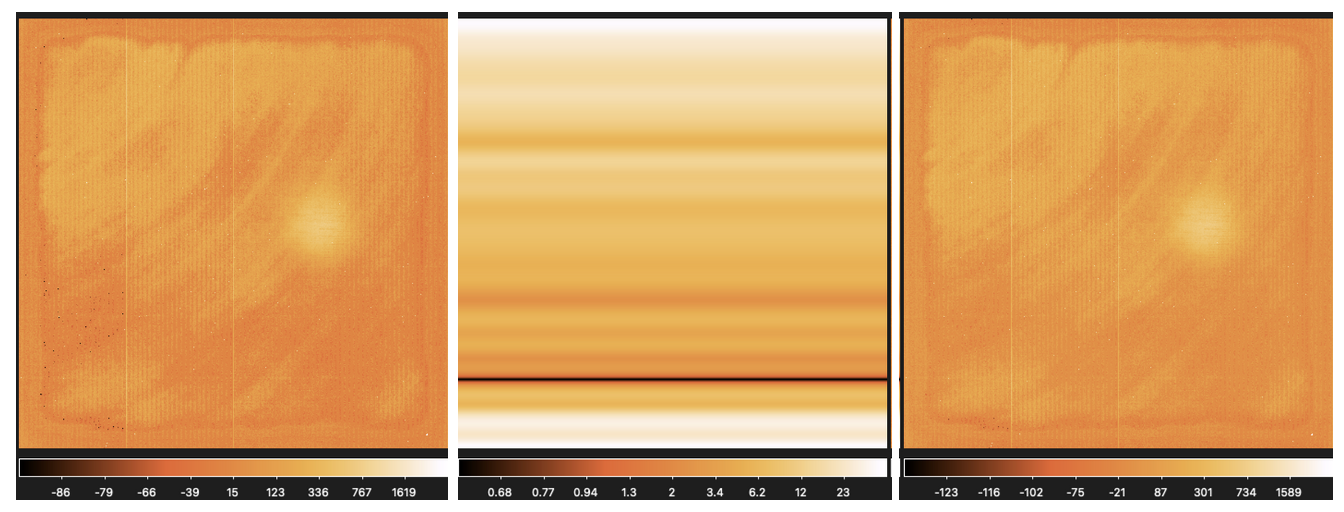}
\end{tabular}
\end{center}
\caption[flicker] 
    { \label{fig:flicker} 
    The ACN corrected read (left), the 1/f noise estimated using the left and right reference pixels (middle), and the 1/f corrected read (right).}  
\end{figure}

The complex FFT of the original reference pixel signal is then multiplied by this calculated optimal filter. The resulting spectrum, now representing the isolated 1/f noise component, is transformed back into the time domain via an inverse FFT. This smoothed noise signal is subsequently subtracted from all pixels in the corresponding science data rows, effectively removing the horizontal striping caused by 1/f noise. The adaptive nature of this algorithm allows it to construct a custom filter based on the unique noise properties of the data, leading to a highly effective correction. Figure \ref{fig:flicker} shows a representative detector read before and after the 1/f correction, along with the derived noise estimate that was subtracted.


\subsection{Linearity Correction}
As an H2RG detector approaches saturation, its response becomes non-linear mainly due to the change in the capacitance of the photodiode within each pixel as it accumulates charge. One has to correct this and make it as linear as possible before doing further data analysis. We employ a pixel-by-pixel non-linearity correction similar to the \textit{HST} WFC3/IR and \textit{JWST} pipeline \cite{Hilbert_2004}. 

\begin{figure} [ht]
\begin{center}
\begin{tabular}{c}  
\includegraphics[width=\textwidth]{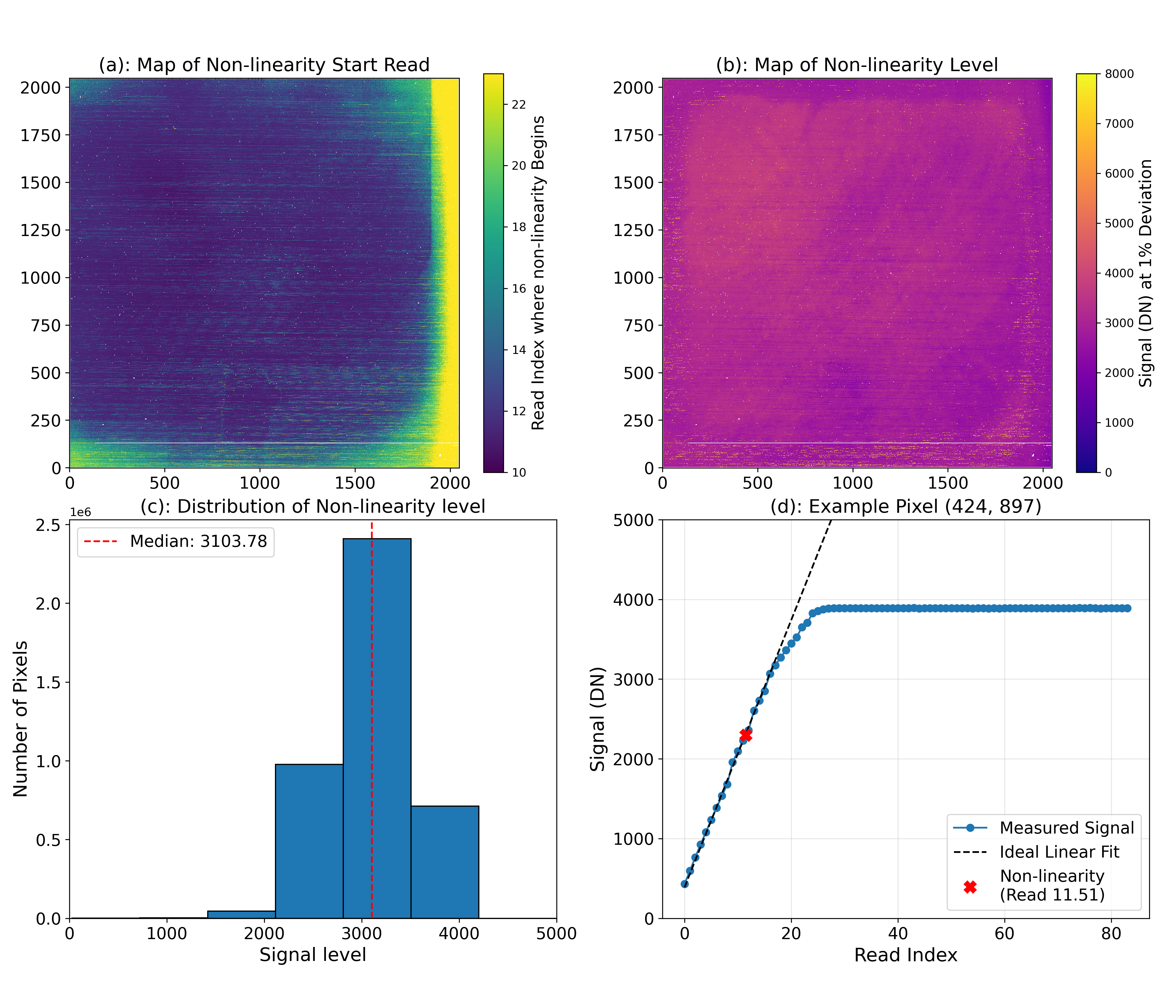}
\end{tabular}
\end{center}
\caption[nonlinearity] 
    { \label{fig:nonlinearity_flat} 
    (a): Map of the starting read for non-linearity, defined as the read where the signal deviates by more than 1\% from the linear model fit. (b): Map of pixel values at the read where the measured signal deviates by more than 1\% from the linear model fit. (c): Histogram of the pixel values from the linearity map in panel (b). (d): Measured signal as a function of read for pixel (424,897) and the best-fit straight line (dashed black line). The threshold read occurs at the 11th read, corresponding pixel value of 3103 DN at a clock rate of 5 MHz.}  
\end{figure}

The characterization of the detector's non-linear response begins with the generation of a high signal-to-noise (S/N) master ramp. This is created by combining multiple flat-field illumination ramps using a sigma-clipping algorithm to reject outliers. From this master ramp, an initial, true linear signal rate is determined for each pixel by fitting a linear model to a carefully selected range of reads. The range begins at the third read to exclude potential electronic instabilities often observed in the initial reads. The upper bound of this linear regime is defined as the read where the measured signal deviates by 1\% from an extrapolation of the initial linear fit through an iterative process. Analysis of the pixel ensemble reveals that this 1\% deviation typically occurs at the eleventh read in the particular detector dark dataset, corresponding to a median signal level of approximately 3103 DN. Therefore, reads 3 through 11 are selected as the definitive range for establishing the true linear signal rate for all pixels. The methodology for selecting this fitting range is illustrated in Figure \ref{fig:nonlinearity_flat}.

To derive the correction function, the true linear signal is plotted against the measured signal. This relationship is fit with a piecewise function consisting of two separate third-order polynomials. The curve is split into two sections at a division point set at 75\% of the maximum linear signal. The coefficients for both polynomials and the corresponding cutoff value for each pixel are then saved to a reference file to be used for the non-linearity correction to all the ramp observations. We will test this approach across all detector clock rates and optimize the threshold values used to derive the polynomial coefficients in future work.

\begin{figure} [ht]
\begin{center}
\begin{tabular}{c}  
\includegraphics[width=0.5\textwidth]{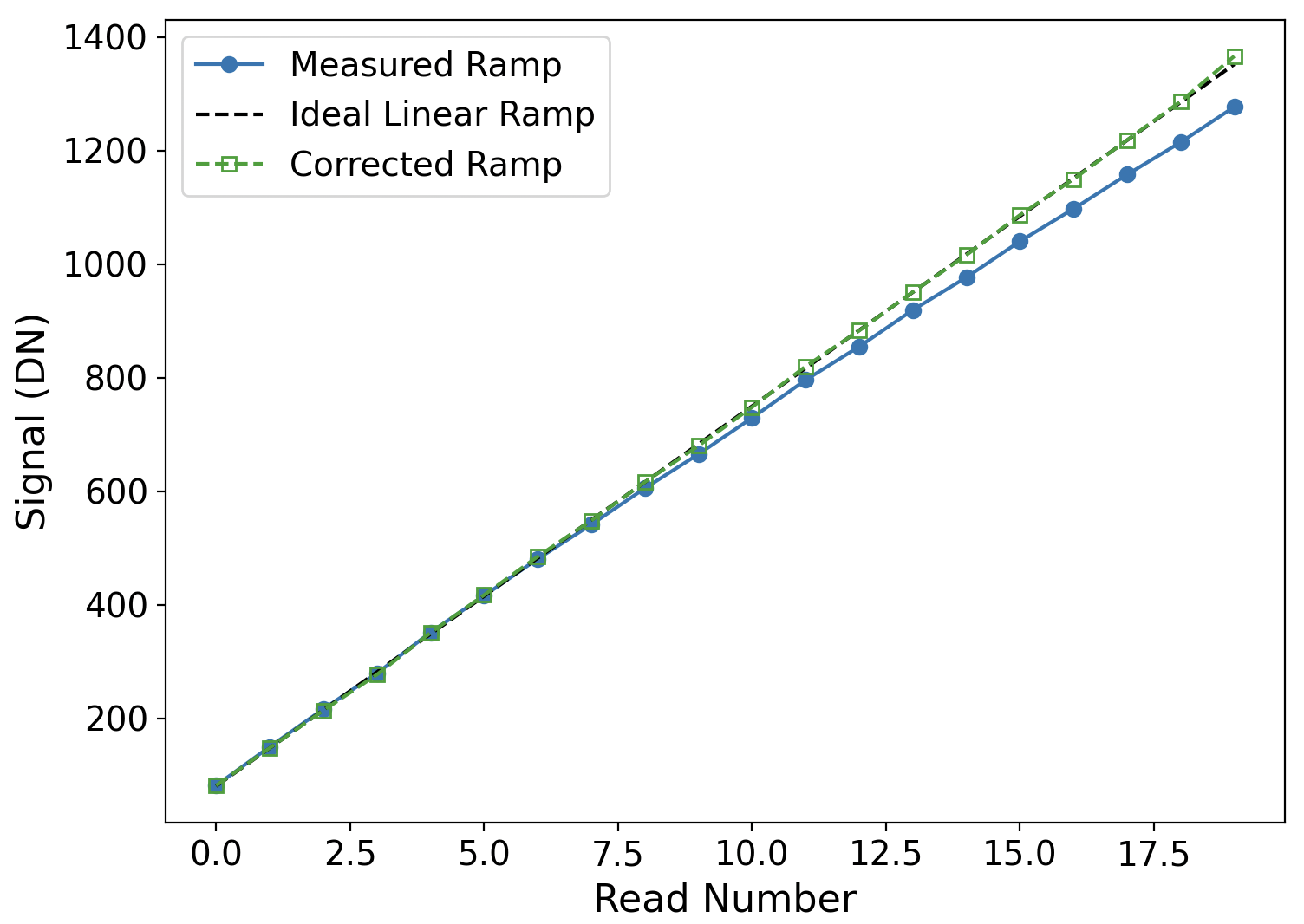}
\end{tabular}
\end{center}
\caption[nonlinearity] 
    { \label{fig:nonlinearity} 
    The measured pixel value from a  five sigma-clipped flat ramps 20 reads (blue line), the best-fit straight line to the data using 1\% cutoff (black dashed line), and the corrected ramp after applying the polynomial coefficients (green line).}  
\end{figure}

A linearity correction is applied to each unsaturated pixel using a third order polynomial equation $$X_{\mathrm{corr}} = a_1 \times X + a_2 \times X^2 + a_3 \times X ^3$$ where $a_1, a_2, a_3$ are the polynomial coefficients, X is the measured count of a single pixel, and $X_{\mathrm{corr}}$  is the corresponding corrected count.  If the pixel value is below the threshold, the first set of coefficients is applied; otherwise, the second set of coefficients is used. Figure \ref{fig:nonlinearity} shows the non-linearity correction applied on a single pixel (10,10) from a set of 20 flat ramps. We observed that the corrected ramp exhibits a standard deviation of 0.42\% for the pixel (10,10).

\subsection{Ramp Fitting (Including Jump Detections)}
\begin{figure} [ht]
\begin{center}
\begin{tabular}{c}  
\includegraphics[height=7cm]{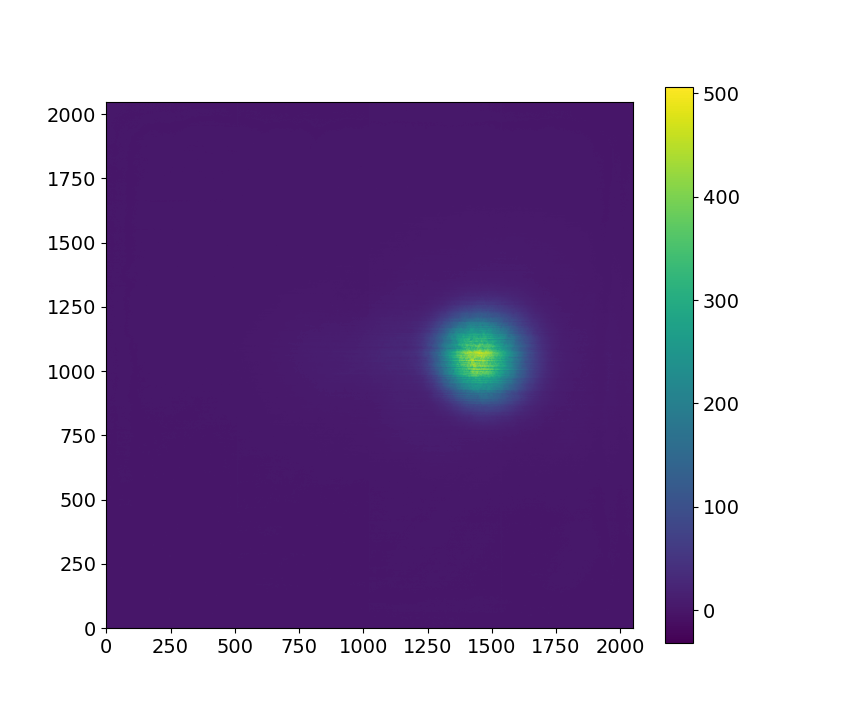}
\end{tabular}
\end{center}
\caption[exposure] 
    { \label{fig:exposure} 
    An exposure or a best-fit slope image using ten reads and the non distructive ramp fitting method by Brandt \cite{Brandt_2024a}.}  
\end{figure}

We determine the count rate of  unsaturated individual pixels using the optimal ramp fitting and jump detection framework developed by Brandt  \cite{Brandt_2024a}. This method applies a Generalized Least Squares (GLS) fit to the differences of consecutive reads. The GLS weighting is determined by a temporal covariance matrix that accurately models the correlated noise sources, including both read noise and the accumulating photon noise. The elements of this matrix representing the variance of each difference and the covariance between adjacent differences are derived analytically from the expected contributions of these noise sources. A complete derivation of the covariance matrix and the fitting procedure is detailed in Brandt \cite{Brandt_2024a}. The use of the covariance matrix within the GLS fit provides optimal data weighting, which in turn removes the statistical bias present in unweighted fits. The integration of likelihood-based jump detection with optimal ramp fitting has been adopted in the official JWST pipeline, where it has demonstrated both robustness and accuracy. Figure \ref{fig:exposure} shows the best-fit slope image from an exposure using ten reads. 

We adopted a likelihood-based method for jump detection and cosmic ray rejection method by Brandt \cite{Brandt_2024b}. The algorithm first establishes a baseline goodness of fit under the null hypothesis that no cosmic ray jump is present. This is achieved by performing a  GLS fit of a single linear ramp to the data and calculating the corresponding chi-squared value. Next, the algorithm tests the alternative hypothesis that a jump occurred. It iterates through every possible interval between reads, treating each as a potential jump location. For each potential location, it fits a "broken line" model composed of an initial flux, a constant slope, and a single jump in amplitude. The chi-squared for this jump model is calculated for each possible location. After testing all possible locations, the algorithm identifies the one that yields the minimum possible chi-squared value for a single-jump model.

This statistical formulation provides a principled means of distinguishing cosmic ray events from noise fluctuations while preserving the integrity of the ramp slope for flux estimation. Benchmarking against conventional thresholding methods demonstrates improved detection efficiency and lower false-positive rates, establishing the likelihood framework as a robust tool for high-precision astronomical measurements.

The GLS method becomes unreliable for fewer than six reads. In that case, we fit the input reads with a first-order polynomial in time, where the slope represents the photon count rate and the intercept captures the initial pixel state. The fit is performed independently for each unsaturated pixel using weighted least squares, with weights that reflect both photon noise, which increases with accumulated charge, and uncorrelated read noise. This procedure averages down read noise across the multiple samples and yields an unbiased estimate of the pixel count rate, provided the detector response is linear and free of discontinuities. The quicklook module also employs this polynomial fitting method for ramp fitting due to its faster execution time.
\subsection{Bad Pixel Correction}
Finally, a bad pixel correction is applied to individual exposures using two approaches. The first is an iterative median filter of kernel size varying from ($3\times3$) to ($11\times11$) with a minimum of $0.45\%$ of good neighbors around each bad pixel. The second method is applying an interpolation using a 2D Gaussian kernel with a standard deviation of three sigma. The details of bad pixel mask generation are explained in Section \ref{bpm}.

\section{SPECTRAL EXTRACTION} \label{spectral_extraction}
The next general step from ramp image to the final 3D IFS datacube is a spectral extraction after flat fielding. The SCALES-DRP implements two main extraction methods, optimal extraction and $\chi^{2}$ extraction. Both methods produce a 3D IFS datacube along with a corresponding flux error cube, with dimensions defined by the number of lenslets in each spatial direction and the number of wavelength bins along the spectral axis. For the imaging channel, the ramp-fitted slope images will be the final output from the SCALES-DRP. 

\subsection{Optimal Extraction} \label{optimal}
The pipeline implements the standard optimal extraction algorithm by Horne \cite{Horne_1986}, applied by several data reduction pipelines in the past \cite{Baranne_1996,Cushing_2004,Brandt_2017}. 
This technique estimates the spectral intensity at each wavelength by weighting the pixels according to the measured line-spread function of the  calibration lamp psflet and the associated pixel specific uncertainties. By using a normalized model of the instrumental point spread function to weight each pixel along the spectral trace, it maximizes the signal-to-noise ratio (S/N) while minimizing the contribution of detector noise and sky background.

We implemented optimal extraction using a linear algebraic rectification matrix generated using the calibration module (Section \ref{calib:module}). Each column in the rectification matrix represents the weight of each pixel at each wavelength, measured from a monochromatic calibration lamp dataset. A total variance map is computed for each pixel by combining contributions from photon noise and read noise, ensuring proper weighting of the data during the optimal extraction process.  

\begin{figure} [ht]
   \begin{center} 
   \includegraphics[height=7cm]{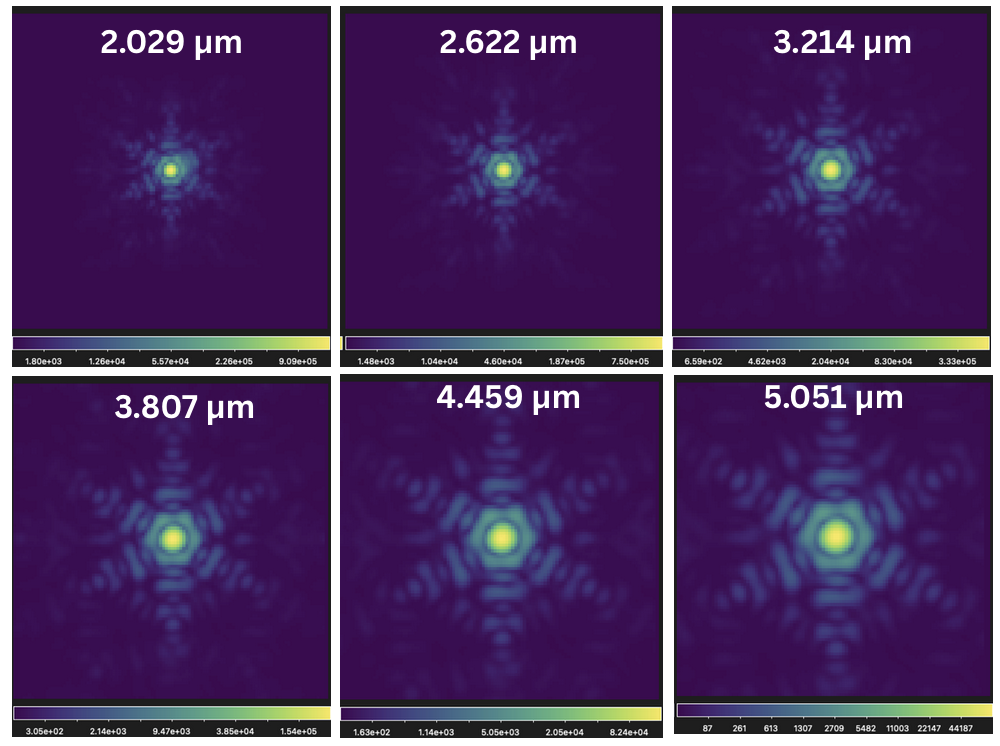}
   \end{center}
   \caption[optimal] 
   { \label{fig:cube} 
    The IFS cube slice at different wavelengths from 2-5 $\mu$m for a simulated input A star data using {\ttfamily scalessim} for low-resolution IFS mode with 54 wavelength bins.}     \end{figure}

The flux amplitudes of the final science datacube are calculated by multiplying the weighted 2D detector image by the rectification matrix, which calculates a weighted sum of flux around individual psflets onto a rectified grid of shape ($\lambda_{bin}$, $X_{lenslets}$, $Y_{lenslets}$) where $\lambda_{bin}$ is the wavelength bins used, $X_{lenslets}$ is the number of lenslets in the x direction, and $Y_{lenslets}$ is the number of lenslets in the Y direction. The output errorcube is the error propagation followed in each step of this optimization process. Currently there is no interpolation performed on the wavelength grid and we are in the process of optimizing  $\lambda_{bin}$ to avoid overestimating or underestimating the measured flux. Due to its rapid performance, executing in tens of milliseconds on a Mac M3 Max with 128 GB of RAM, the optimal extraction algorithm is used to generate the quicklook IFS data cubes. Figure \ref{fig:cube} shows the IFS data cube slices from simulated observations at different wavelength covering 2-5 $\mu$m.

\subsection{$\chi^2$ Extraction}
The $\chi^{2}$ based spectral extraction differs from optimal extraction by fitting the entire two-dimensional data at predefined psflet positions using a rectification matrix of shape (4194304, $X_{lenslet} \times Y_{lenslet} \times \lambda_{bin}  $ ). Each column in the rectification matrix represents the weight of each pixel at each wavelength, measured from a calibration unit dataset. The one-dimensional optimal flux from each psflet is extracted using a forward-modeling technique that formulates the extraction as a linear inverse problem, $R \times A = d$, where R is the rectification matrix, A is the best-fit flux for each psflet as a function of wavelength, and $d$ is the observed 2D detector image. The optimal solution for $A$ is found by minimizing the chi-squared statistic, which represents the variance-weighted sum of squared residuals between the forward-model and the input data. The weights are derived from both detector read noise and signal-dependent photon noise. To ensure a physically plausible result, the optimization is subject to two constraints: a non-negativity constraint on the flux ($A$ $\geq$ 0), and a dynamic upper bound derived from optimal extraction (refer Section \ref{optimal}). This bounded, weighted, non-negative least-squares (BWNNLS) system is solved efficiently using a trust-region reflective algorithm, as implemented in {\ttfamily scipy.optimize.lsq-linear}.

\begin{figure} [ht]
   \begin{center}
   \begin{tabular}{c} 
   \includegraphics[width=\textwidth]{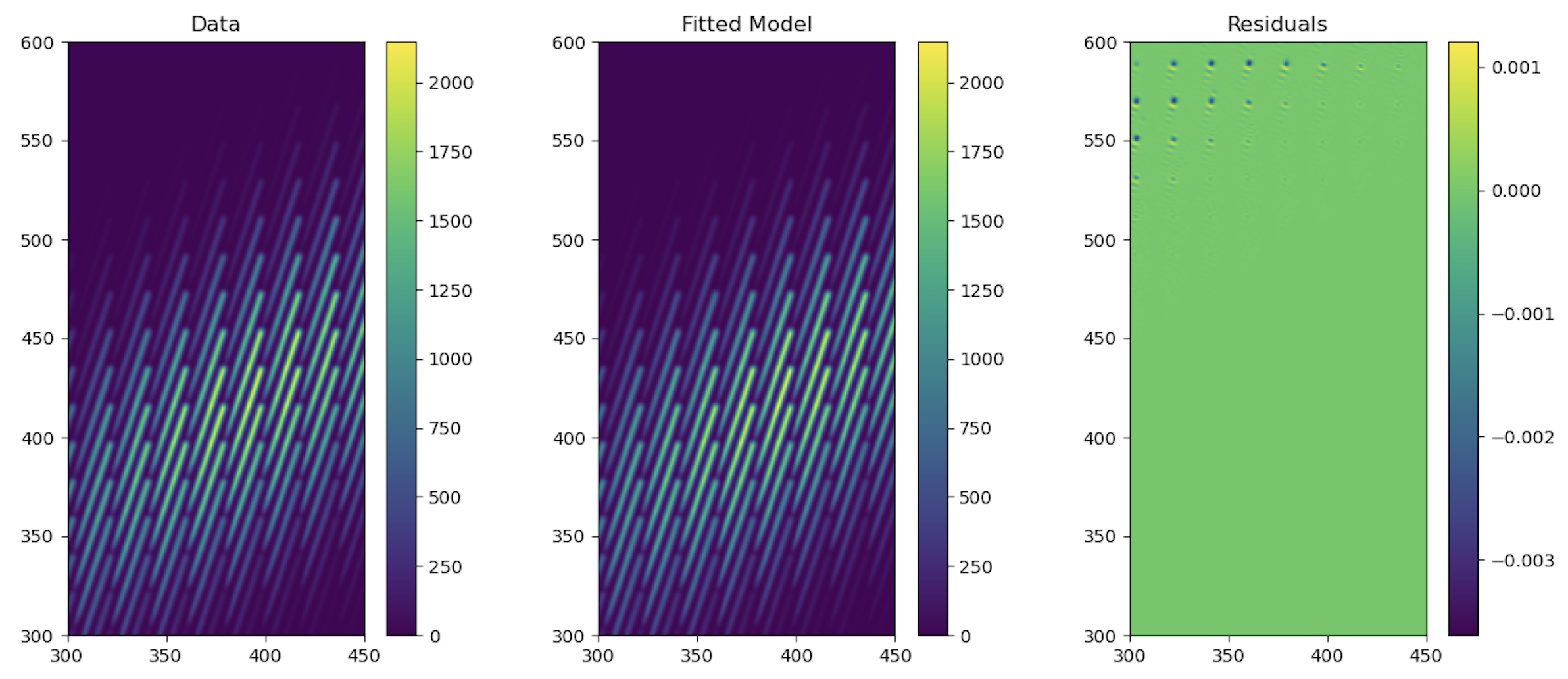}
   \end{tabular}
   \end{center}
   \caption[chi] 
   { \label{fig:chi} 
    A simulated input data using {\ttfamily scalessim} for low-resolution IFS mode with 54 wavelength bins (left), the best-fit $\chi^{2}$ extracted model (middle), and residual (right) for a subset of the detector pixels.}  
   \end{figure}

Setting the upper bound value to the {\ttfamily lsq-linear} using the optimal extraction output drastically improved the execution time from the order of hours to minutes on a M3 Max with 128GB RAM. This approach is slightly different from CHARIS \cite{Brandt_2017} and GPI \cite{Berdeu_2020} where individual microspectra are cropped on a predefined box shape and a least square fitting is performed using the corresponding normalized calibration psflets. While a separation of 6 pixels between adjacent microspectra helps to mitigate optical crosstalk on the SCALES detector, the $\chi^2$ fit is capable of removing the remaining crosstalk. Since the fit is performed over the full two-dimensional microspectrum, it also naturally allows for simultaneous modeling and subtraction of a uniform, undispersed background. Figure \ref{fig:chi} shows a simulated low-resolution SCALES datacube using {\ttfamily scalessim} and the best-fit $\chi^2$ model to the data.

\section{Conclusion}
SCALES is a thermal infrared integral field spectrograph and diffraction-limited imager under development for the W. M. Keck Observatory. We detail the design and implementation of the SCALES Data Reduction Pipeline (SCALES-DRP), which processes raw detector reads into science-ready data products for its two operational modes: integral field spectroscopy (IFS) and imaging. The pipeline first applies a suite of detector-level corrections, including reference pixel subtraction, 1/f noise filtering, non-linearity correction, and bad pixel masking. The count rate for each pixel is then determined via optimal ramp fitting. For IFS data, the pipeline performs flat-fielding before employing two complementary spectral extraction algorithms: a variance-weighted optimal extraction and a forward-modeling extraction based on a bounded, non-negative least-squares ($\chi^2$) minimization. These methods enable the robust reconstruction of 3D data cubes with propagated flux uncertainties. For the imaging mode, the final data product is the flat-fielded slope image after ramp fitting.

The DRP operates in two distinct modes. A quicklook mode, accessible via graphical user interfaces (GUIs), provides on-site data quality assessment and real-time feedback to observers, ensuring efficient use of telescope time. In its science-grade processing mode, the pipeline delivers high-fidelity, archival-quality data products suitable for a wide range of astrophysical investigations, from the detection and characterization of exoplanets to studies of circumstellar disks, Solar System objects, and extragalactic targets. A dedicated calibration module generates the necessary reference files and instrumental models used throughout both the quicklook and science-grade processing workflows.

As SCALES proceeds toward commissioning, continued DRP development will focus on algorithm optimization, refinement of calibration strategies, and the incorporation of user feedback. Data from the upcoming laboratory cool-down tests will be used to validate and optimize the DRP algorithms for both the IFS and imager. The pipeline is expected to be fully operational for SCALES’ first light on the Keck II telescope, anticipated in early 2026.

\acknowledgments 
Major grants to the SCALES project include NSF Grant 2216481 as well as grants from the Heising-Simons Foundation, the Mt. Cuba Astronomical Foundation, and the Alfred P. Sloan Foundation.  We are also grateful to the Robinson family and other private supporters, without whom this work would not be possible.

\normalsize
\setlength{\baselineskip}{0.8\baselineskip}
\bibliography{report} 

\begin{thebibliography}{10}

\bibitem{Scott_haka}
Lilley, S.~J., Wizinowich, P., Steiner, J., Wetherell, E., Thorne, J., Marin, E., Kassis, M., Guthery, C., Delorme, J., Bouchez, A., Roberts, M., Jovanovic, N., Sanchez, D., and Hinz, P., ``{Keck AO high order wavefront sensing and control: opto-mechanical design},'' in [{\em Adaptive Optics Systems IX}{\nolinebreak\hspace{0.1em}]},  Jackson, K.~J., Schmidt, D., and Vernet, E., eds.,  {\bf 13097},  130977B, International Society for Optics and Photonics, SPIE (2024).

\bibitem{sallum_2023}
Sallum, S., Skemer, A., Stelter, D., Banyal, R., Batalha, N., Batalha, N., Blake, G., Brandt, T., Briesemeister, Z., de~Kleer, K., de~Pater, I., Desai, A., Eisner, J., fai Fong, W., Greene, T., Honda, M., Jensen-Clem, R., Kain, I., Kilpatrick, C., Kupke, R., Lach, M., Liu, M.~C., Macintosh, B., Martinez, R.~A., Mawet, D., Miles, B., Morley, C., Powell, D., Sethuram, R., Sheehan, P., Spilker, J., Stone, J., Surya, A., Thirupathi, S., Unni, A., Wagner, K., and Zhou, Y., ``{The slicer combined with array of lenslets for exoplanet spectroscopy (SCALES): driving science cases and expected outcomes},'' in [{\em Techniques and Instrumentation for Detection of Exoplanets XI}{\nolinebreak\hspace{0.1em}]},  Ruane, G.~J., ed.,  {\bf 12680},  1268003, International Society for Optics and Photonics, SPIE (2023).

\bibitem{surya_2024}
{Surya}, A., {Kupke}, R., {Stelter}, D., {Hinz}, P., {Sallum}, S., {Hasan}, A., {Skemer}, A., {Thirupathi}, S., {Banyal}, R., and {Unni}, A., ``{Performance analysis of SCALES final optical design: end to end modeling},'' in [{\em Ground-based and Airborne Instrumentation for Astronomy X}{\nolinebreak\hspace{0.1em}]},  {Bryant}, J.~J., {Motohara}, K., and {Vernet}, J. R.~D., eds., {\em Society of Photo-Optical Instrumentation Engineers (SPIE) Conference Series} {\bf 13096},  130966W (July 2024).

\bibitem{Banyal_2022}
{Banyal}, R.~K., {Hasan}, A., {Kupke}, R., {Varshney}, H.~M., {Prakaesh}, A., {Sivarani}, T., {Skemer}, A.~J., {MacDonald}, N., {Sallum}, S., {Deich}, W., {Fitzgerald}, M.~P., {Govinda}, K.~V., {Ratliff}, C., {Sethuram}, R., {Stelter}, D., {Surya}, A., and {Wang}, E., ``{Design of an IR imaging channel for the Keck Observatory SCALES instrument},'' in [{\em Advances in Optical and Mechanical Technologies for Telescopes and Instrumentation}{\nolinebreak\hspace{0.1em}]},  {\em Society of Photo-Optical Instrumentation Engineers (SPIE) Conference Series} {\bf 12188},  121881U (Aug. 2022).

\bibitem{Rauscher_2014}
{Rauscher}, B.~J., ``{New and Better H2RG Detectors for the JWST Near Infrared Spectrograph},'' in [{\em American Astronomical Society Meeting Abstracts \#223}{\nolinebreak\hspace{0.1em}]},  {\em American Astronomical Society Meeting Abstracts} {\bf 223},  149.39 (Jan. 2014).

\bibitem{Deno_2020}
{Deno Stelter}, R., {Skemer}, A.~J., {Sallum}, S., {Kupke}, R., {Hinz}, P., {Mawet}, D., {Jensen-Clem}, R., {Ratliffe}, C., {MacDonald}, N., {Deich}, W., {Kruglikov}, G., {Kassis}, M., {Lyke}, J., {Briesemeister}, Z., {Miles}, B., {Gerard}, B., {Fitzgerald}, M., {Brandt}, T., and {Marois}, C., ``{Update on the Preliminary Design of SCALES: the Santa Cruz Array of Lenslets for Exoplanet Spectroscopy},'' {\em arXiv e-prints} ,  arXiv:2012.09098 (Dec. 2020).

\bibitem{Skemer_2022}
{Skemer}, A.~J., {Stelter}, R.~D., {Sallum}, S., {MacDonald}, N., {Kupke}, R., {Ratliff}, C., {Banyal}, R., {Hasan}, A., {Varshney}, H.~M., {Surya}, A., {Prakaesh}, A., {Thirupathi}, S., {Sethuram}, R., {K.~V.}, G., {Fitzgerald}, M.~P., {Wang}, E., {Kassis}, M., {Absil}, O., {Alvarez}, C., {Batalha}, N., {Boucher}, M.-A., {Bourgenot}, C., {Brandt}, T., {Briesemeister}, Z., {de Kleer}, K., {de Pater}, I., {Deich}, W., {Divakar}, D., {Filion}, G., {Gauvin}, {\'E}., {Gonzales}, M., {Greene}, T., {Hinz}, P., {Jensen-Clem}, R., {Johnson}, C., {Kain}, I., {Kruglikov}, G., {Lach}, M., {Landry}, J.-T., {Li}, J., {Liu}, M.~C., {Lyke}, J., {Magnone}, K., {Marin}, E., {Martin}, E., {Martinez}, R., {Mawet}, D., {Miles}, B., {Sandford}, D., {Sheehan}, P., {Sohn}, J.~M., and {Stone}, J., ``{Design of SCALES: a 2-5 micron coronagraphic integral field spectrograph for Keck Observatory},'' in [{\em Ground-based and Airborne Instrumentation for Astronomy IX}{\nolinebreak\hspace{0.1em}]},  {Evans}, C.~J., {Bryant}, J.~J., and
  {Motohara}, K., eds., {\em Society of Photo-Optical Instrumentation Engineers (SPIE) Conference Series} {\bf 12184},  121840I (Aug. 2022).

\bibitem{Kupke_2022}
{Kupke}, R., {Stelter}, R.~D., {Hasan}, A., {Surya}, A., {Kain}, I., {Briesemeister}, Z., {Li}, J., {Hinz}, P., {Skemer}, A., {Gerard}, B., {Dillon}, D., and {Ratliff}, C., ``{SCALES on Keck: optical design},'' in [{\em Ground-based and Airborne Instrumentation for Astronomy IX}{\nolinebreak\hspace{0.1em}]},  {Evans}, C.~J., {Bryant}, J.~J., and {Motohara}, K., eds., {\em Society of Photo-Optical Instrumentation Engineers (SPIE) Conference Series} {\bf 12184},  121844A (Aug. 2022).

\bibitem{Macintosh_2018}
{Macintosh}, B., {Chilcote}, J.~K., {Bailey}, V.~P., {de Rosa}, R., {Nielsen}, E., {Norton}, A., {Poyneer}, L., {Wang}, J., {Ruffio}, J.~B., {Graham}, J.~R., {Marois}, C., {Savransky}, D., and {Veran}, J.-P., ``{The Gemini Planet Imager: looking back over five years and forward to the future},'' in [{\em Adaptive Optics Systems VI}{\nolinebreak\hspace{0.1em}]},  {Close}, L.~M., {Schreiber}, L., and {Schmidt}, D., eds., {\em Society of Photo-Optical Instrumentation Engineers (SPIE) Conference Series} {\bf 10703},  107030K (July 2018).

\bibitem{Beuzit_2006}
{Beuzit}, J.~L., {Feldt}, M., {Dohlen}, K., {Mouillet}, D., {Puget}, P., {Antichi}, J., {Baruffolo}, A., {Baudoz}, P., {Berton}, A., {Boccaletti}, A., {Carbillet}, M., {Charton}, J., {Claudi}, R., {Downing}, M., {Feautrier}, P., {Fedrigo}, E., {Fusco}, T., {Gratton}, R., {Hubin}, N., {Kasper}, M., {Langlois}, M., {Moutou}, C., {Mugnier}, L., {Pragt}, J., {Rabou}, P., {Saisse}, M., {Schmid}, H.~M., {Stadler}, E., {Turrato}, M., {Udry}, S., {Waters}, R., and {Wildi}, F., ``{SPHERE: A 'Planet Finder' Instrument for the VLT},'' {\em The Messenger}~{\bf 125},  29 (Sept. 2006).

\bibitem{Groff_2015}
{Groff}, T.~D., {Kasdin}, N.~J., {Limbach}, M.~A., {Galvin}, M., {Carr}, M.~A., {Knapp}, G., {Brandt}, T., {Loomis}, C., {Jarosik}, N., {Mede}, K., {McElwain}, M.~W., {Leviton}, D.~B., {Miller}, K.~H., {Quijada}, M.~A., {Guyon}, O., {Jovanovic}, N., {Takato}, N., and {Hayashi}, M., ``{The CHARIS IFS for high contrast imaging at Subaru},'' in [{\em Techniques and Instrumentation for Detection of Exoplanets VII}{\nolinebreak\hspace{0.1em}]},  {Shaklan}, S., ed., {\em Society of Photo-Optical Instrumentation Engineers (SPIE) Conference Series} {\bf 9605},  96051C (Sept. 2015).

\bibitem{Horne_1986}
{Horne}, K., ``{An optimal extraction algorithm for CCD spectroscopy.},'' {\em pasp}~{\bf 98},  609--617 (June 1986).

\bibitem{Brandt_2017}
{Brandt}, T.~D., {Rizzo}, M., {Groff}, T., {Chilcote}, J., {Greco}, J.~P., {Kasdin}, N.~J., {Limbach}, M.~A., {Galvin}, M., {Loomis}, C., {Knapp}, G., {McElwain}, M.~W., {Jovanovic}, N., {Currie}, T., {Mede}, K., {Tamura}, M., {Takato}, N., and {Hayashi}, M., ``{Data reduction pipeline for the CHARIS integral-field spectrograph I: detector readout calibration and data cube extraction},'' {\em Journal of Astronomical Telescopes, Instruments, and Systems}~{\bf 3},  048002 (Oct. 2017).

\bibitem{Briesemeister_2020}
{Briesemeister}, Z., {Sallum}, S., {Skemer}, A., {Stelter}, R.~D., {Hinz}, P., and {Brandt}, T., ``{End-to-end simulation of the SCALES integral field spectrograph},'' in [{\em Ground-based and Airborne Instrumentation for Astronomy VIII}{\nolinebreak\hspace{0.1em}]},  {Evans}, C.~J., {Bryant}, J.~J., and {Motohara}, K., eds., {\em Society of Photo-Optical Instrumentation Engineers (SPIE) Conference Series} {\bf 11447},  114474Z (Dec. 2020).

\bibitem{Lyke_2017}
{Lyke}, J., {Do}, T., {Boehle}, A., {Campbell}, R., {Chappell}, S., {Fitzgerald}, M., {Gasawy}, T., {Iserlohe}, C., {Krabbe}, A., {Larkin}, J., {Lockhart}, K., {Lu}, J., {Mieda}, E., {McElwain}, M., {Perrin}, M., {Rudy}, A., {Sitarski}, B., {Vayner}, A., {Walth}, G., {Weiss}, J., {Wizanski}, T., and {Wright}, S., ``{OSIRIS Toolbox: OH-Suppressing InfraRed Imaging Spectrograph pipeline}.'' Astrophysics Source Code Library, record ascl:1710.021 (Oct. 2017).

\bibitem{Lockhart_2019}
{Lockhart}, K.~E., {Do}, T., {Larkin}, J.~E., {Boehle}, A., {Campbell}, R.~D., {Chappell}, S., {Chu}, D., {Ciurlo}, A., {Cosens}, M., {Fitzgerald}, M.~P., {Ghez}, A., {Lu}, J.~R., {Lyke}, J.~E., {Mieda}, E., {Rudy}, A.~R., {Vayner}, A., {Walth}, G., and {Wright}, S.~A., ``{Characterizing and Improving the Data Reduction Pipeline for the Keck OSIRIS Integral Field Spectrograph},'' {\em AJ}~{\bf 157},  75 (Feb. 2019).

\bibitem{Rauscher_2015}
Rauscher, B.~J., ``Teledyne h1rg, h2rg, and h4rg noise generator,'' {\em Publications of the Astronomical Society of the Pacific}~{\bf 127},  1144 (nov 2015).

\bibitem{Kosarev_1983}
Kosarev, E.~L. and Pantos, E., ``Optimal smoothing of 'noisy' data by fast fourier transform,'' {\em Journal of Physics E: Scientific Instruments}~{\bf 16},  537 (jun 1983).

\bibitem{Hilbert_2004}
{Hilbert}, B., ``{Non-Linearity Correction Algorithm for the WFC3 IR Channel}.'' Instrument Science Report WFC3-2004-06, 9 pages (May 2004).

\bibitem{Brandt_2024a}
{Brandt}, T.~D., ``{Optimal Fitting and Debiasing for Detectors Read Out Up-the-Ramp},'' {\em pasp}~{\bf 136},  045004 (Apr. 2024).

\bibitem{Brandt_2024b}
{Brandt}, T.~D., ``{Likelihood-based Jump Detection and Cosmic Ray Rejection for Detectors Read Out Up-the-ramp},'' {\em pasp}~{\bf 136},  045005 (Apr. 2024).

\bibitem{Baranne_1996}
{Baranne}, A., {Queloz}, D., {Mayor}, M., {Adrianzyk}, G., {Knispel}, G., {Kohler}, D., {Lacroix}, D., {Meunier}, J.~P., {Rimbaud}, G., and {Vin}, A., ``{ELODIE: A spectrograph for accurate radial velocity measurements.},'' {\em aaps}~{\bf 119},  373--390 (Oct. 1996).

\bibitem{Cushing_2004}
Cushing, M.~C., Vacca, W.~D., and Rayner, J.~T., ``Spextool: A spectral extraction package for spex, a 0.8–5.5 micron cross‐dispersed spectrograph,'' {\em Publications of the Astronomical Society of the Pacific}~{\bf 116},  362 (apr 2004).

\bibitem{Berdeu_2020}
{Berdeu}, A., {Soulez}, F., {Denis}, L., {Langlois}, M., and {Thi{\'e}baut}, {\'E}., ``{PIC: a data reduction algorithm for integral field spectrographs. Application to the SPHERE instrument},'' {\em aap}~{\bf 635},  A90 (Mar. 2020).

\end{thebibliography}
\bibliographystyle{spiebib} 

\end{document}